\documentclass[sigconf]{acmart}
\usepackage{graphicx} 
\usepackage{ulem}
\renewcommand\footnotetextcopyrightpermission[1]{}

\acmConference[ ]{ }{ }{ }

\makeatletter
\def\@copyrightspace{}
\makeatother

\title{Machine Learning for Consistency Violation Faults Analysis}

\author{Kamal Giri}
\affiliation{%
  \institution{University of Wyoming}
  \city{Laramie}
  \country{USA}}
\email{kgiri@uwyo.edu}

\author{Amit Garu}
\affiliation{%
  \institution{University of Wyoming}
  \city{Laramie}
  \country{USA}}
\email{agaru@uwyo.edu}
\begin{document}

\begin{abstract}
Distributed systems frequently encounter consistency violation faults (cvfs), where nodes operate on outdated or inaccurate data, adversely affecting convergence and overall system performance. This study presents a machine learning-based approach for analyzing the impact of cvfs, using Dijkstra’s Token Ring problem as a case study. By computing program transition ranks and their corresponding effects, the proposed method quantifies the influence of cvfs on system behavior. To address the state space explosion encountered in larger graphs, two models are implemented: a Feedforward Neural Network (FNN) and a distributed neural network leveraging TensorFlow’s \texttt{tf.distribute} API. These models are trained on datasets generated from smaller graphs (3 to 10 nodes) to predict parameters essential for determining rank effects. Experimental results demonstrate promising performance, with a test loss of 4.39 and a mean absolute error of 1.5. Although distributed training on a CPU did not yield significant speed improvements over a single-device setup, the findings suggest that scalability could be enhanced through the use of advanced hardware accelerators such as GPUs or TPUs.
\end{abstract}

\maketitle

\section{Motivation}

With the increased utilization of distributed systems, there has been a growing emphasis on improving fault tolerance, particularly against consistency violation faults (cvfs). These faults occur when neighboring nodes operate on incorrect or stale data, leading to actions based on outdated information. Although prior studies \cite{NGK2023ICDCN} have shown that tolerating cvfs has minimal impact on program performance, analyzing the overhead introduced by cvfs becomes increasingly important when handling large program state spaces.

This work investigates the effect of cvfs by computing the program transitions of two self-stabilizing algorithms and evaluating the associated rank effects. The rank effect is defined as the difference between the destination and source configuration ranks, serving as a proxy for measuring the impact of cvfs on convergence time.

\begin{figure}[ht]
    \centering
    \includegraphics[width=0.4\textwidth]{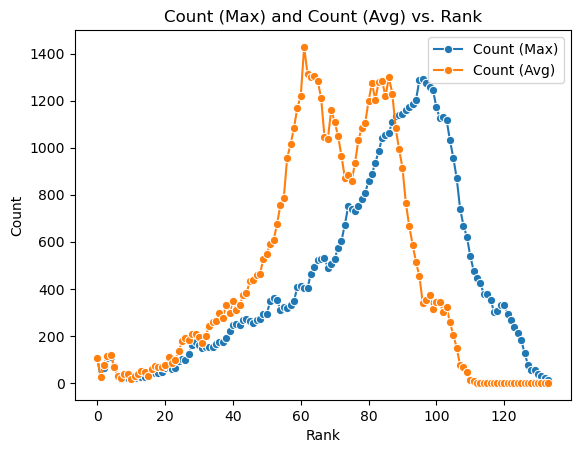}
    \caption{Rank vs. program states count for the Token Ring problem with 10 nodes}
    \Description{Rank vs. program states count for the Token Ring problem with 10 nodes}
    \label{fig:node10}
\end{figure}
\begin{figure}[ht]
    \centering
    \includegraphics[width=0.4\textwidth]{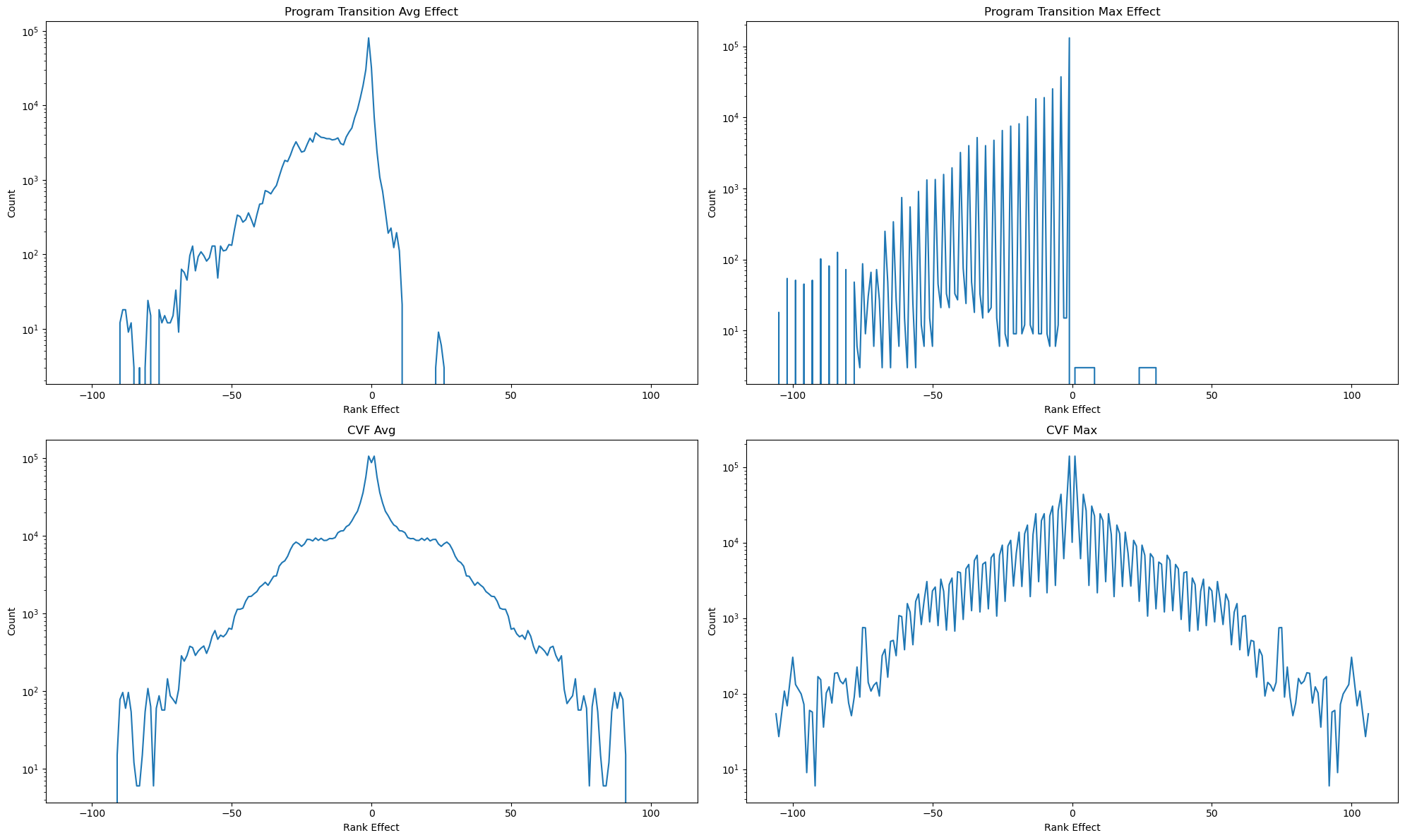}
    \caption{Program transition and CVF rank effect plot for Dijkstra’s Token Ring problem with 10 nodes}
    \Description{Program transition and CVF rank effect plot for Dijkstra’s Token Ring problem with 10 nodes}
    \label{fig:node10a}
\end{figure}

The primary objective of this analysis is to quantify the effect of cvfs on arbitrary graphs by calculating the program transition rank and CVF-induced rank effects in Dijkstra's Token Ring system, focusing initially on a 10-node configuration. As shown in Figure~\ref{fig:node10}, the highest concentration of states occurs at ranks between 60 and 80. Figure~\ref{fig:node10a} illustrates the rank effects associated with program transitions, cvfs entering a state (\(cvfs~in\)), and cvfs exiting a state (\(cvfs~out\)).

While the rank effect was computable for 10-node systems, increasing the number of nodes results in a state space explosion, rendering exhaustive computation infeasible. To overcome this limitation, a machine learning-based approach is proposed, employing a Feedforward Neural Network (FNN) and a Distributed Neural Network to predict rank effects in larger graph configurations.

\section{Introduction}

The volume of training data required for larger models, such as neural networks, increases exponentially with the number of parameters. As the demand for processing large-scale data continues to exceed the growth in computational power, distributing the machine learning workload across multiple machines has become increasingly essential \cite{survey}. When analyzing consistency violation faults (cvfs) using Dijkstra's Token Ring problem, the state space expands rapidly with the number of nodes, rendering traditional analytical approaches inefficient for systems with high node counts.

To address this challenge, the present work investigates both conventional and distributed machine learning techniques. Specifically, it explores the application of a Feedforward Neural Network (FNN) and a distributed neural network trained using TensorFlow's \texttt{tf.distribute} API. These models are implemented and evaluated in order to mitigate the limitations caused by state space explosion and to enable predictive analysis of rank effects in larger graph configurations.

The broader objective of this research is to integrate and apply core concepts from machine learning, distributed computing, and parallel optimization algorithms. Key areas of focus include the architecture of distributed training systems, inter-process communication mechanisms, the use of parameter servers, and the implementation of both synchronous and asynchronous stochastic gradient descent (SGD) algorithms.

Rather than addressing these elements in isolation, the study demonstrates how these components can be effectively combined to solve a practical problem.

\section{Background and Related Work}
\subsection{Machine Learning }
Machine learning algorithms learn to make decisions and predictions based on data. They are typically categorized based on three key characteristics: feedback, which allows the model to gradually improve its performance; purpose, which defines the specific task or problem the model is designed to address; and method, which refers to the approach the model uses to improve itself based on new input data and enhance its accuracy \cite{survey}.
\subsection{Distributed Machine Learning}
There are two fundamentally different and complementary ways of accelerating workloads: adding more resources to a single machine (vertical scaling) and adding more nodes to the system (horizontal scaling) \cite{survey}. Horizontal scaling, also known as distributed computing, forms the foundation for distributed machine learning (DML). DML refers to the use of multiple nodes or machines to train machine learning models in parallel. This approach is particularly effective when the dataset is too large to fit on a single machine, enabling benefits such as speedup, scalability, fault tolerance, parallelization, improved model accuracy, and cost efficiency \cite{survey}.
DML typically involves two main approaches: Data Parallelism and Model Parallelism. Since we are using TensorFlow, which primarily supports data parallelism, this paper will focus on data parallelism.
Some key components of DML include:

\subsubsection{Data Parallelism}
In data parallelism, the training data is split across multiple devices or nodes, with each device computing gradients for its subset of the data. These gradients are then aggregated and used to update the global model parameters.
\begin{figure}[ht]
    \centering
    \includegraphics[width=0.5\linewidth]{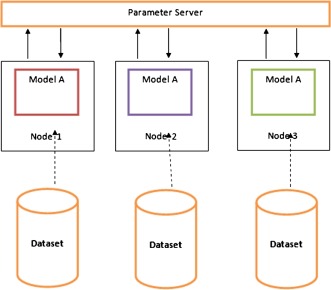}
    \caption{Data parallelism \cite{Parallelism_fig}}
    \Description{Data parallelism}
    \label{fig:enter-label}
\end{figure}

\subsubsection{Gradient Aggregation and Synchronization}
In distributed training, gradients computed on different devices or workers must be aggregated to produce a global gradient update. In synchronous aggregation, all workers compute their local gradients and these gradients are aggregated.In asynchronous aggregation, workers independently compute and push their gradients to a parameter server. The server updates the model parameters without waiting for all workers.
Synchronization refers to coordinating the timing of updates among the workers to ensure a consistent model state \cite{tensorflow}.

\subsubsection{Parameter Server}
The Parameter Server framework is a scalable and efficient system for distributed machine learning, designed to handle the growing complexity of large-scale datasets and models. It distributes data and computational workloads across worker nodes, while server nodes manage globally shared parameters using dense or sparse vectors and matrices. Key features include asynchronous communication to reduce network overhead, flexible consistency models to balance efficiency and convergence, elastic scalability for dynamic node addition, and robust fault tolerance with rapid recovery mechanisms\cite{Parameter}.

\subsection{TensorFlow}
TensorFlow is a machine learning system that operates at large scale and in heterogeneous environments, which uses dataflow graphs to represent computation, shared state, and the operations that mutate that state.TensorFlow uses a single dataflow graph to represent all computation and state in a machine learning algorithm, including the individual mathematical operations, the parameters and their update rules, and the input preprocessing. Dataflow makes the communication between subcomputations explicit, and therefore
makes it easy to execute independent computations in parallel, and partition the computation across multiple distributed devices \cite{Tensorflow_paper}.
TensorFlow uses tf.distribute API to train the data in multiple nodes, and we will be using this API to train this model as well.
There are different strategies in the tf.distribute which essentially tell what kind of model we want to use. 
Some Strategies:
\begin{list}{--}{}
    \item tf.distribute.MirroredStrategy
    \item tf.distribute.MultiWorkerMirroredStrategy
    \item tf.distribute.TPUStrategy
    \item tf.distribute.CentralStorageStrategy
    \item tf.distribute.ParameterServerStrategy
    \item AllReduce for Gradient Aggregation:
\end{list}
as well as what frameworks/tools like NVIDIA NCCL as well as communication protocols \cite{tensorflow}.

\subsection{Consistency Violation Faults (cvfs)}

As distributed systems continue to grow in scale and complexity, there is an increasing need to enhance their resilience to faults. One notable class of faults is consistency violation faults (cvfs), which occur when neighboring nodes operate on outdated or inaccurate data, leading to erroneous behavior \cite{NGK2023ICDCN}.

Previous work \cite{NGK2023ICDCN} investigated the impact of cvfs using the concept of rank effect. However, computing the rank effect for graphs with a large number of nodes leads to state space explosion, making exhaustive analysis infeasible. To address this challenge, the present study proposes a machine learning-based solution using both Feedforward and Distributed Neural Networks. These models are trained on datasets derived from lower-node graphs to enable prediction of rank effects in larger configurations.

The rank effect is defined as:

\begin{equation}
    \text{rank effect} = \text{rank(dest)} - \text{rank(start)}
\end{equation}

Once the rank effects for program transitions and cvfs (in and out) are calculated, additional parameters \( M \) and \( Ar \) are derived based on the following definitions:

\begin{itemize}
    \item \( L \): Total path length
    \item \( C \): Path count
    \item \( A = \frac{L}{C} \): Average path length
    \item \( Ar = \lceil A \rceil \): Rounded average path length
    \item \( M = \text{max} + 1 \): Max adjustment parameter
\end{itemize}

After computing the values of \( M \) and \( Ar \), the analysis proceeds by counting how many program transitions or CVF events correspond to each value pair. The resulting distributions are then visualized as plots, showing the frequency of events against corresponding rank effect metrics \( Ar \) and \( M \).

\section{Methodology}
\subsection{Dataset}

One of the most critical and challenging aspects of this work is the construction of a suitable dataset. As previously discussed, calculating the program transition rank effect, cvfs-in rank effect, and cvfs-out rank effect requires obtaining the program transition rank, which assigns a rank to every possible configuration of the ring graph using five key values: \(L\), \(C\), \(A\), \(Ar\), and \(M\). 

All invariant configurations are assigned default values of \(L = 0\), \(C = 1\), \(A = 0\), \(Ar = 0\), and \(M = 0\). Variant configurations differ in these values depending on their distance from the invariants. Among the five, only three values are independent since \(A = L/C\) and \(Ar = \lceil A \rceil\). Therefore, if the values of \(L\), \(C\), and \(M\) are known, the rank effect can be determined and used for model training and evaluation.

To reduce complexity, the dataset in this study focuses on predicting only the value of \(Ar\). Although earlier versions of the model attempted to predict both \(Ar\) and \(M\), doing so significantly increased the difficulty of the learning task.

The final dataset is structured as a vector \(v\), defined as follows:

\begin{itemize}
    \item \( v[0:N] \): state configuration
    \item \( v[N] \): number of nodes \(N\)
    \item \( v[N+1 : I_N - 1] \): placeholder values (0 or 1) for padding
    \item \( v[I_N] \): target value (e.g., \(M\))
\end{itemize}

with the following definitions:

\begin{itemize}
    \item \(N\): total number of nodes in the graph
    \item \(I_N\): total number of input neurons
    \item \(v[0:I_N-1]\): feature vector (independent variables)
    \item \(v[I_N]\): label (dependent variable)
\end{itemize}

The value of \(I_N\) also defines the maximum supported graph size for prediction. For instance, if \(I_N = 15\), the model can be used to predict \(Ar\) for graphs with up to 15 nodes. One challenge with this design is the use of placeholder values to accommodate dynamic input sizes, which may introduce redundancy or sparsity in the feature space.

\subsection{Model Architecture}

Once the dataset was prepared, the next step involved designing and training the predictive models. Two types of neural networks were implemented for this task: a Feedforward Neural Network (FNN) and a distributed neural network utilizing TensorFlow's \texttt{tf.distribute} API.

\subsubsection{Feedforward Neural Network (FNN)}

The Feedforward Neural Network (FNN) serves as the baseline model in this study. Its architecture and training configuration are as follows:

\begin{itemize}
    \item \textbf{Learning Rate}: Set to 0.001 and optimized using the Adam optimizer to balance convergence speed and training stability.
    \item \textbf{Activation Functions}: ReLU is used for the hidden layers, while a linear activation is applied to the output layer for regression.
    \item \textbf{Optimizer}: Adam optimizer with adaptive learning rate.
    \item \textbf{Architecture}: Three hidden layers with 128, 64, and 64 neurons, respectively, and a single output neuron.
    \item \textbf{Regularization}: Dropout (rate of 0.2) applied after the first hidden layer; batch normalization applied after the second hidden layer.
    \item \textbf{Epochs and Batch Size}: Trained for 300 epochs with a batch size of 32.
    \item \textbf{Loss Function}: Mean Squared Error (MSE), suitable for regression tasks.
    \item \textbf{Evaluation Metric}: Mean Absolute Error (MAE) is used to assess prediction accuracy.
\end{itemize}

\subsubsection{Distributed Neural Network}

The second model utilizes distributed training through TensorFlow’s \texttt{tf.distribute.MirroredStrategy}. While the architecture mirrors that of the FNN, the key distinction lies in the training strategy, which replicates the model across multiple devices to enable synchronous gradient updates.

\begin{itemize}
    \item \textbf{Learning Rate}: 0.001, optimized using the Adam optimizer.
    \item \textbf{Activation Functions}: ReLU for hidden layers and linear activation for the output layer.
    \item \textbf{Optimizer}: Adam optimizer with adaptive learning rate.
    \item \textbf{Network Architecture}:
    \begin{itemize}
        \item Three hidden layers with 128, 64, and 64 neurons.
        \item One output layer with a single regression neuron.
    \end{itemize}
    \item \textbf{Regularization}: Dropout (rate of 0.2) applied to the first hidden layer; batch normalization after the second hidden layer.
    \item \textbf{Epochs and Batch Size}: 200 epochs with a batch size of 64, distributed across available CPU cores.
    \item \textbf{Loss Function}: Mean Squared Error (MSE).
    \item \textbf{Evaluation Metric}: Mean Absolute Error (MAE).
    \item \textbf{Distributed Training}: Training is executed synchronously across multiple devices using \texttt{MirroredStrategy}.
\end{itemize}

\section{Results and Discussion}

Code and datasets used for this study are available at: \newline \href{https://github.com/kgiri2000/Cvf-Analysis-Distributed-Machine-Learning}{github.com/kgiri2000/Cvf-Analysis}

Considerable effort was dedicated to hyperparameter tuning, as achieving optimal results proved to be particularly challenging. Certain components, such as regularization layers, were removed during experimentation to simplify the model and improve convergence. Although the final performance did not completely align with initial expectations, the results show promising potential and indicate clear directions for future refinement.

The dataset was constructed for graphs with node counts ranging from 3 to 10, and the model was trained on this data. Following training, the model was evaluated on graphs with 10 and 11 nodes to predict the corresponding \(M\) values. The evaluation yielded a test loss of 4.39 (Mean Squared Error) and a test MAE of 1.5 (Mean Absolute Error), suggesting that, on average, the model's predictions deviate by 1.5 units from the actual values—a reasonable level of accuracy given the complexity of the task.

\begin{figure}[ht]
    \centering
    \includegraphics[width=1\linewidth]{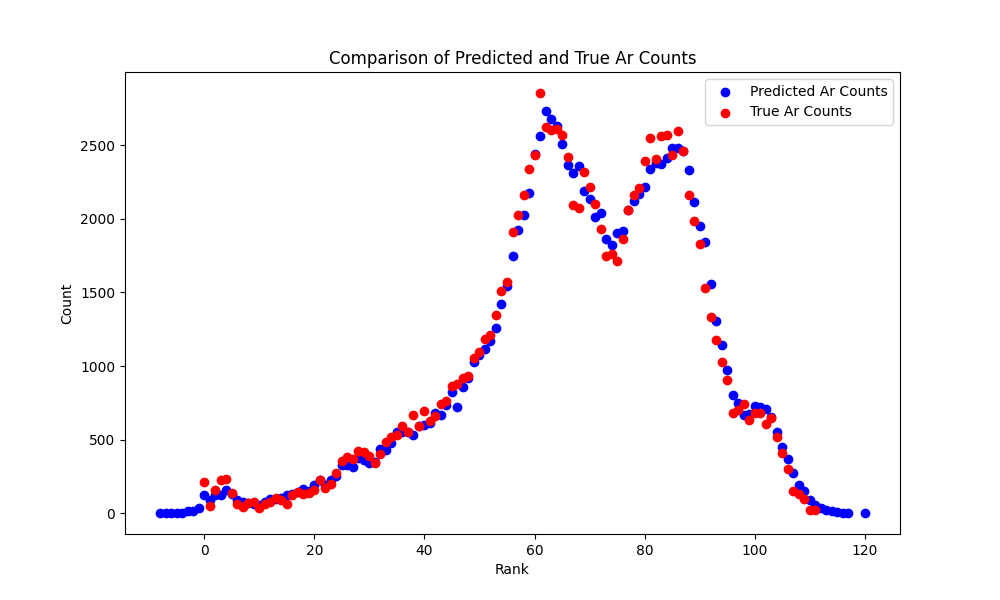}
    \caption{Rank vs. count for graph with 10 nodes}
    \Description{Rank vs. count for graph with 10 nodes}
    \label{p10}
\end{figure}

\begin{figure}[ht]
    \centering
    \includegraphics[width=1\linewidth]{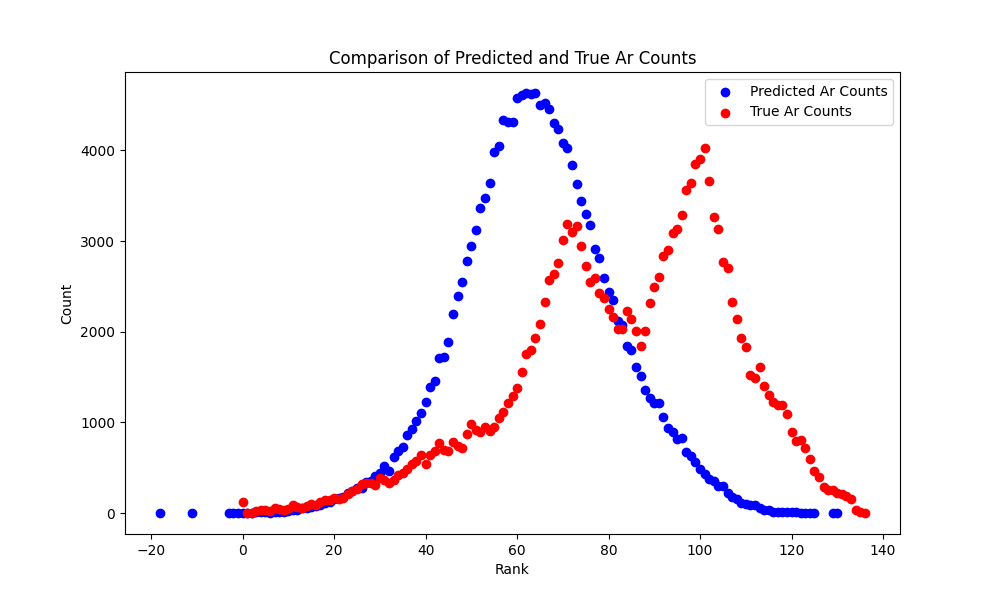}
    \caption{Rank vs. count for graph with 11 nodes}
    \Description{Rank vs. count for graph with 11 nodes}
    \label{p11}
\end{figure}

As shown in Figure~\ref{p10}, the model was able to replicate the general distribution of rank counts for a graph with 10 nodes. However, this does not imply that the predicted \(M\) values were correct for all individual configurations; it simply indicates that the overall count distribution followed the expected trend. The plotted rank counts are cumulative and do not verify the correctness of each individual prediction.

Figure~\ref{p11} presents the full set of predicted \(M\) values for a graph with 11 nodes. These predictions were aggregated to generate the corresponding rank count graph. The results exhibit recognizable patterns and trends, indicating that the model is learning meaningful representations. Nonetheless, improvements such as additional hyperparameter tuning, stronger regularization, or alternative training methods could further enhance prediction accuracy.

The distributed training implementation was executed using TensorFlow with \texttt{tf.distribute.MirroredStrategy} on CPU cores. However, the execution time did not significantly differ from the single-device configuration—approximately two minutes for 30 epochs—suggesting that further emphasis on distributed training is not yet justified under the current hardware constraints. Performance gains may become more apparent when utilizing GPUs or TPUs for larger datasets or more complex models.

\section{Conclusion and Future Work}
This project addressed the challenge of analyzing consistency violation faults (cvfs) in distributed graph algorithms, particularly Dijkstra's Token Ring Problem. By leveraging machine learning models, including a Feedforward Neural Network (FNN) and TensorFlow’s \texttt{tf.distribute} API, the approach mitigated the state space explosion problem for larger node graphs.

The results showed promising trends, with reasonable accuracy for smaller graphs, though challenges such as overfitting and computational inefficiencies in distributed training were identified. Future work will focus on using the predicted M values to calculate rank effects (program transitions, cvfs in, and cvfs out) for larger graphs, along with optimizing distributed training on GPUs or TPUs for better scalability. This project provided valuable insights into neural networks and distributed machine learning, laying the groundwork for further advancements in the field.

\section{Acknowledgments}
This research was supported by the National Science Foundation (NSF) through the Research Experiences for Undergraduates (REU) program at the University of Wyoming.

The project builds upon prior work presented in \cite{NGK2023ICDCN} and was conducted during the Summer of 2024. The authors thank Professor Duong Ngoc Nguyen for providing the opportunity to pursue this research and for offering continuous guidance throughout the project. Appreciation is also extended to Amit Garu for his assistance in implementing Dijkstra’s algorithm and for his valuable contributions during the development of this work.

\bibliographystyle{ACM-Reference-Format}
\bibliography{References}

\end{document}